\documentclass[showpacs,twocolumn,aps,preprintnumbers,letterpaper]{revtex4}
\usepackage{amsmath,amssymb}
\usepackage{epsfig}
\usepackage{graphicx}
\usepackage{amsmath}
\usepackage{amsfonts}
\usepackage{epstopdf}

\newcommand{\be}{\begin{equation}}
\newcommand{\ee}{\end{equation}}
\newcommand{\bear}{\begin{eqnarray}}
\newcommand{\eear}{\end{eqnarray}}
\newcommand{\ba}{\begin{array}}
\newcommand{\ea}{\end{array}}

\def\be{\begin{eqnarray}}
\def\ee{\end{eqnarray}}

\def\roughly#1{\mathrel{\raise.3ex\hbox{$#1$\kern-.75em%
\lower1ex\hbox{$\sim$}}}}

\begin{document}

\title{Probing Wilson Loops in the QCD Instanton Vacuum}

\author{Yizhuang Liu}
\email{yizhuang.liu@stonybrook.edu}
\affiliation{Department of Physics and Astronomy, Stony Brook University, Stony Brook, New York 11794-3800, USA}

\author{Ismail Zahed}
\email{ismail.zahed@stonybrook.edu}
\affiliation{Department of Physics and Astronomy, Stony Brook University, Stony Brook, New York 11794-3800, USA}

\date{\today}

\begin{abstract}
We discuss the quark and gluon condensates in the presence of a rectangular Wilson loop using the QCD instanton vacuum with 
three light dynamical quarks. The scalar quark condensate is found to decrease while the gluon condensate to increase. We also
derive the static potential between two QCD dipoles and show that it is attractive but short ranged at large distances. Its relevance to static QCD
string interactions is discussed.
\end{abstract}

\maketitle

\setcounter{footnote}{0}


\section{Introduction}
QCD is the fundamental theory of strong interactions. It has proven challenging in the infrared
as its fundamental constituents are confined. In this regime, the quarks and gluons
interact strongly and form strongly interacting hadrons. The instanton approach
to the QCD vacuum allows for a semi-classical description of the QCD vacuum that is well
supported by lattice simulations~\cite{Schafer:1996wv,Diakonov:2002fq,Nowak:1996aj}
(and references therein) .
It captures the essentials of the spontaneous breaking of
chiral symmetry, accounts for the breaking of conformal symmetry 
and provides a  mechanism for a large
mass for the pseudo-scalar singlet through mixing with the fluctuations in the topological charge.

\begin{figure}[htb]
\centerline{
\includegraphics[width=6cm]{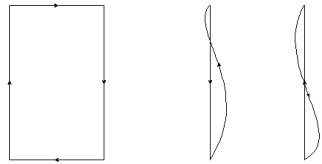}}
\caption{Large Wilson loop:  unscreened in YM (left) and screened in QCD (right). See text.}
\label{WILLABX}
\end{figure}

The instanton approach to the Yang-Mills vacuum does not provide a mechanism for confinement.
However, in the presence of light quarks the QCD string breaks asymptotically owing to screening by
the fundamental charges.  In the presence of light quarks a large Wilson loop traced by a
heavy quark pair at a separation $a$ is screened by the formation of a pair of heavy-light
mesons as shown in Fig.~\ref{WILLABX} (right). Interactions of light-light and heavy-light quarks can be 
addressed simultaneously in the QCD instanton vacuum~\cite{Chernyshev:1994zm,Chernyshev:1995gj}.

Several lattice simulations have shown that the scalar quark condensate depletes near a heavy quark 
sourced by a Polyakov line~\cite{Buerger:1993bq} or a static quark-antiquark string spanned by a Wilson line~\cite{Iritani:2014jqa}. The
squared topological density was also found to decrease near a heavy quark source~\cite{Faber:1993fj}.
In this paper we probe the quark and gluon clouds around a small and large  Wilson loop in the QCD instanton
vacuum with heavy and light quarks. Many of our results are amenable to a quantitative comparison 
with QCD lattice simulations in the unquenched approximation using Wilson loops.

In section
2 we briefly review the salient features of the QCD instanton
vacuum and summarize the relevant quark and
gluon correlations. The effective interactions between
light and heavy quarks are also briefly discussed. In section
3 we analyze the scalar condensates  around a small
Wilson loop in the one-instanton approximation. In section 4 we derive explicit relations for
the scalar, pseudo-scalar quark and gluon condensates in
the presence of a large Wilson loop in the screened QCD
instanton vacuum. In section 5 we discuss the dominant
contributions
to the static dipole-dipole interaction both
for small and large dipole sizes. Our conclusions are in
section 6.

\section{QCD Instanton Vacuum}

Throughout this section we will follow the notations and the results of the general $N_f$ bosonization scheme developed
in~\cite{Kacir:1996qn}. The original $N_f=2$ bosonization scheme was developed by Diakonov and Petrov
and is summarized in~\cite{Diakonov:2002fq}.  The reader interested in more details regarding the derivations
should consult these two references for completeness. Throughout, we will use Euclidean notations unless specified otherwise.

\subsection{Light Quarks}

One of the key aspect of the QCD instanton vacuum is that light quarks acquire a
momentum dependent constituent quark mass by rescattering
through randomly distributed instanton and anti-instanton configurations. Specifically and for massless light quarks
~\cite{Diakonov:2002fq,Kacir:1996qn}

\be
S(x)=\left<T^*  q(x)q^\dagger (0)\right> =\int \frac{d^4k}{(2\pi)^4}\frac {e^{ik\cdot x}}{\gamma \cdot p-iM(k)}
\ee
with  the momentum dependent constituent quark mass

\be
M(k)=\frac{\lambda n}{N_c}k^2\left(\pi {\bar  \rho}^2\frac d{dx} (I_0K_0-I_1K_1)\right)^2
\label{MK}
\ee
\newline
where $x=k\bar\rho/2$. Note that the quark zero mode is expressed in terms of Bessel functions in Fourier space. The instanton density fixes the gluon condensate in the QCD instanton vacumm $\left<F^2\right>=32\pi^2 n$. 
The inverse mass parameter $\lambda\approx 0.34\,{\rm fm}$ follows
by self-consistency. Typically $n\approx 1/{\rm fm}^4$  and $\bar\rho\approx 1/3\,{\rm fm}$. As a result, the vacuum breaks spontaneously
chiral symmetry and develops a chiral condensate. For a single flavor species

\be
\left<\bar q q\right>=-4N_c\int \frac{d^4k}{(2\pi)^4}\frac{M(k)}{k^2+M(k)^2}
\ee
with $\bar q\equiv (iq)^\dagger$ the quark field in Minkowski space. Throughout the Euclidean notation
will be used unless specified otherwise. A good approximation to $S(x)$ at large $|x|$ follows from using
$M_0=M(k\approx 0)\approx 300-400$ MeV, with

\be
S(x)\approx \frac{iM_0^2}{4\pi^2}\left(\frac{\gamma \cdot x}{x^2}K_2(M_0 x)+\frac 1x  K_1(M_0 x)\right)
\ee

Typical scalar quark correlations in the QCD instanton vacuum are captured by

\begin{equation}
C(x)\equiv \left< T^\star q^{\dagger} q(x) q^{\dagger}q(0)\right>=C^0(x)+C^1(x)
\label{C}
\end{equation}
where the  first term is the connected contribution in Fig.1 (left)

\be
C^0(x)={\rm Tr}\left|\gamma_5S(x)\right|^2
\ee
which is dominant for small $|x|$.
The second contribution stems from the disconnected contribution of Fig.1 (right).
 In the mean-field approximation its Fourier transform is~\cite{Diakonov:2002fq,Kacir:1996qn}

\be
C^1(p)=2N_c \frac{R^2(p)}{\Delta_+(p)}
\label{RSIGMA}
\ee
For small 4-momentum

\be
\Delta_{+}(p)=\frac{f_+^2}{4N_c}\left(M_+^2+p^2+{\cal O}(m^2,p^4)\right)
\ee
For $N_f=3$  light flavor masses $m =5$ MeV: $f_\sigma\equiv f_+=110$ MeV and $m_\sigma\equiv
M_+=640$ MeV.
This exchanged meson is identified with the light scalar meson with $m_\sigma\equiv M_+<2M_0$,
below the $q\bar q$ threshold.  The residue at the pole is~\cite{Diakonov:2002fq,Kacir:1996qn}

\be
R(0)= \int \frac{d^4k}{(2\pi)^2}\frac{2}{M(k)} \frac{k^2/M(k)^2-1}{(1+k^2/M(k)^2)^2}
\ee
so that
\be
C^1(x)\approx \frac{4N_c^2R^2(0)}{f_\sigma^2} \frac{m_\sigma}{4\pi^2|x|}K_1(m_\sigma |x|)
\ee
which is dominant for large $|x|$.

\begin{figure}[htb]
\centerline{
\includegraphics[width=6cm]{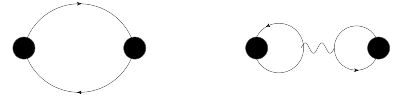}}
\caption{Meson correlator: connected (left) and disconnected (right). See text.}
\label{WILLAB}
\end{figure}

The pseudo-scalar correlator follows from similar arguments with

\begin{equation}
C_5(x)\equiv \left<T^*q^\dagger \gamma_5 q(x) q^\dagger \gamma_5 q(0)\right>=C_5^0(x)+C_5^1(x)
\label{C5}
\end{equation}
with the connected part

\be
C_5^0(x)={\rm Tr}\left|S(x)\right|^2
\ee
and the disconnected part~\cite{Diakonov:2002fq,Kacir:1996qn}

\be
C_5^1 (p)=N_c\frac{R^2_5(p)}{\Delta_-(p)+3\chi_{\star}/2N_c}
\ee
Here $\chi_\star=n\approx 1/{\rm fm}^{4}$ is the quenched topological susceptibility, and
\be
R_5(0) = -2\int \frac{d^4k}{(2\pi)^2}\frac{M(k)}{M(k)^2+k^2}
\ee
Again, for small 4-momentum

\be
\Delta_{-}(p)=\frac{f_-^2}{4N_c}\left(M_-^2+p^2+{\cal O}(m^2,p^4)\right)
\ee
For $N_f=3$  light flavor masses $m=5$ MeV:  $f_\pi\equiv f_-=88$ MeV and $m_\pi\equiv M_-=160$ MeV,
are the pseudo-scalar decay constant and mass respectively. $m_\pi\equiv M_-$ obeys the GOR relation~\cite{GellMann:1968rz} and vanishes in the chiral limit.

(14) captures the soft pion theorem for the pseudoscalar coupling to a constituent quark with fixed flavor
$g_\pi\equiv \left<\bar q q\right>/f_\pi =2N_cR_5(0)/f_\pi$, in terms of which (13) reads
\be
C^1_5(p)\approx \frac{g^2_\pi}{p^2+m_\pi^2+2N_f\chi_\star/f_\pi^2}
\ee
which exhibits explicitly the $\eta'$ pole with $m^2_{\eta'}=m_\pi^2+2N_f\chi_\star/f_\pi^2$. 
Typically $\left<\bar q q\right>\approx (-271\,{\rm MeV})^3$ from the GOR
relation with the pion parameters quoted, so that $g^2_\pi\approx (475\,{\rm MeV})^4$
and $m_{\eta^\prime}\approx 1125\,{\rm MeV}$.
Similarly, we may identify the scalar coupling  to a constituent quark of fixed flavor as $g_\sigma=2\sqrt{2}N_cR(0)/f_\sigma$, in terms of which (7) takes the canonical form
\be
C^1(p) \approx \frac{g^2_\sigma}{p^2+m_\sigma^2}
\ee
with the $\sigma$ pole explicit. Chiral symmetry implies $g_\pi \approx g_\sigma$ between the scalar and
pseudo-scalar coupling.

\subsection{Gluons}

The gluonic correlations in the QCD instanton vacuum mix with the fermionic correlations as shown in Fig.3.
Throughout we will  define ${\rm Tr}(gF)^2\equiv F^2$ unless noted otherwise. With this in mind,
the scalar gluon correlation function is~\cite{Kacir:1996qn}


\begin{eqnarray}
&&\left<T^*F^2 (x)  F^2 (0)\right>_c=\nonumber\\
&&\int \frac{d^4p}{(2\pi)^4} e^{ip\cdot x}\left(\frac{1}{\sigma^2_{\star}}-\frac{N_f}{2N_c\Delta_+(p)}\right)^{-1}
\end{eqnarray}
The ultra-local contribution is the  is the quenched gluon compressibility 
$\sigma_\star^2={12n/11N_c}\approx 0.36/{\rm fm}^4$.
Thus

\begin{eqnarray}
&&\left<T^*F^2 (x)F^2 (0)\right>_c=\nonumber\\
&&\sigma^2_{\star}\delta^4(x)+\frac{N_f\sigma_\star^4}{f_\sigma^2}\frac{m_s}{2\pi^2|x|}K_1(m_s |x|)
\label{GLUON}
\end{eqnarray}
with $m_s^2 =m_\sigma^2-2N_f\sigma^2_{\star}/f_\sigma^2$ which is lighter than $m_\sigma$ due to the
fluctuations in the instanton density as captured by the quenched gluon compressibility. 
For the typical
vacuum parameters quoted earlier, $m_s\approx 349\,{\rm MeV}$ which is in the range of the
constituent quark mass $M_0\approx 300-400\, {\rm MeV}$ and  below 
the sigma meson $m_\sigma\approx 640\,{\rm MeV}$.

\begin{figure}[htb]
\centerline{
\includegraphics[width=6cm]{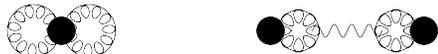}}
\caption{Gluon correlator: connected and ultra-local (left) and disconnected (right). See text.}
\label{WILLAB}
\end{figure}

The pseudo-scalar gluon correlation function~\cite{Diakonov:2002fq,Kacir:1996qn}

\begin{eqnarray}
&&\left<T^*F \tilde F (x)F \tilde F (0)\right>=\nonumber\\
&&\int \frac{d^4p}{(2\pi)^4} e^{ip\cdot x}\left(\frac{1}{\chi_{\star}}+\frac{N_f}{2N_c\Delta_-(p)}\right)^{-1}
\label{FFTFFT}
\end{eqnarray}
 simplifies to

\begin{eqnarray}
&&\left<T^*F \tilde F (x)F \tilde F (0)\right>=\nonumber\\
&&\chi_{\star}\delta^4(x)-\frac{N_f\chi_\star^2}{f_\pi^2}\frac{m_{\eta^\prime}}{2\pi^2|x|}K_1(m_{\eta^\prime}|x|)
\end{eqnarray}

We note that in the quenched approximation 
($N_f=0$) both the scalar and pseudo-scalar glueballs are infinitely heavy in leading order in the QCD
instanton vacuum. The fluctuations in the instanton and anti-instanton
numbers cause the scalar and pseudo-scalar glueballs to be lighter through mixing. Specifically:
$m_s\approx m_\sigma-N_f/N_c^2$ (scalar glueball) and $m_{ps}\equiv m_{\eta^\prime}\approx m_\pi+N_f/N_c$
(pseudo-scalar glueball). The $1/N_c$ shifts are sub-leading but tied to the QCD anomalies: the
former to the scale anomaly and the latter to the U(1) axial
anomaly. Overall, they maybe affected by higher order and non-anomalous $1/N_c$ corrections 
neglected in the bosonization analysis in~\cite{Kacir:1996qn}.

\subsection{Heavy and Light Quarks}

In the instanton vacuum heavy and light  quarks interact. The effective vertices for heavy-light interactions
were discussed in~\cite{Chernyshev:1994zm,Chernyshev:1995gj}. For 2 light flavors in the presence of
a heavy quark $Q$, the induced interaction is typically
of the form~\cite{Chernyshev:1994zm,Chernyshev:1995gj}

\begin{eqnarray}
{\cal L}_{qqQ}\approx -i\kappa_{qqQ}\,\,
{Q^\dagger}\frac {(1+i\gamma^4)}2\,Q\,\left({\rm det}q^\dagger_L q_R+{\rm det}q^\dagger_R q_L\right)
\label{qqQ}
\end{eqnarray}
A Fierz rearrangement of (\ref{qqQ}) reads

\begin{eqnarray}
&&
-i\kappa_{qqQ}\,\,
{Q^\dagger}\frac {(1+i\gamma^4)}8\,Q\  \\
&&\times (u^\dagger u d^\dagger d+ u^\dagger \gamma_5 u d^\dagger \gamma_5 d +u^\dagger \gamma^{\nu\mu}ud^\dagger\gamma^{\nu\mu}d)\nonumber
\label{uud}
\end{eqnarray}
which will be used below.
(\ref{qqQ}) is to be compared to the standard t$^\prime$ Hooft interaction~\cite{'tHooft:1976fv}

\be
{\cal L}_{qqq}\approx i\kappa_{qqq}\,\,\left({\rm det}q^\dagger_Lq_R+{\rm det}q^\dagger_Rq_L\right)
\ee
for $N_f=3$ light flavors. Typically: $\kappa_{qqQ}/\kappa_{qqq}\approx 0.08$~\cite{Chernyshev:1994zm}.
For the parameters of the QCD instanton vacuum $\kappa_{qqQ}\approx 0.012\,{\rm fm}^{5}$ with a constituent quark mass
$M_0\approx 400$ MeV. The interaction  (\ref{qqQ}) induces bound scalar, pseudo-scalar and tensor
states with a typical size at the origin~\cite{Chernyshev:1995gj}

\be
|\phi_{qQ}(0)|^2\equiv \frac 1{a_{qQ}^3}\approx \left(\frac{2\pi^2f_{qQ}^2}{N_c}\right)^{3/2}
\ee
For a D-meson $f_D\leq 290\,{\rm MeV}$ so that $a_{qQ}\approx 0.26\,{\rm fm}$.

\section{Small Wilson loop in the One-Instanton approximation}


We first assess the effects of a single instanton and anti-instanton on a small Wilson loop of size a.
The case of a large Wilson loop in the presence of many instantons and anti-instantons will be considered next.
To estimate the scalar condensate near the center $x$
of the Wilson-loop we use the OPE expansion in the form

\be
\frac {W}{\left<W\right>}=1- \frac{2a^4 \pi^2}{N_c}\,F^2(x)+{\cal O}(a^6)
\ee
The leading operator contribution to the scalar condensate in a Wilson-loop follows from the mixed operator $\left< q^\dagger q\,F^2\right>$.  In the one-instanton approximation and for arbitrary flavors $N_f$

\begin{eqnarray}
\left< q^\dagger q(x)\,F^2(x)\right>\approx \frac{384 N_f}{\pi^{2}}
\int \frac{d\mu_{B+F}}{im}\,\frac{\rho^{6}}{ (|x-x_o|^2+\rho^2)^7}
\end{eqnarray}
with the one-loop density~\cite{Bernard:1979qt}

\begin{eqnarray}
\frac{d\mu_{B+F}}{d^4x_0d\rho}&&=\frac{2^{4N+2} \pi^{4N-2}}{(N-1)!(N-2)!}\frac{1}{\pi^{2(N_f-1)}} \nonumber\\
&&\times \left(\frac{\rho M_{PV}}{g}\right)^{4N}\left(\frac m{M_{PV}}\right)^{N_f}e^{-\frac{8\pi^2}{g^2}}
\label{DENS}
\end{eqnarray}
with $M_{PV}$ a Pauli-Villars UV regulator.
The quark condensate in the one-instanton approximation is

\begin{eqnarray}
\left< q^\dagger q (x) \right>\approx\frac{4N_f}{\pi^{2}}
\int \frac{d\mu_{B+F}}{im} \frac{\rho^{2}}{ (|x-x_o|^2+\rho^2)^3}
\end{eqnarray}
Thus

\begin{eqnarray}
\frac{\left< q^\dagger q(x) W\right>_c}{\left< q^\dagger q\right>\left<W\right>}\approx-\frac{64\pi^2}{5N_c}\frac{a^4}{\bar{\rho}^{4}}
\end{eqnarray}
where we have fixed $\rho\approx \bar \rho$ in the instanton density.

The scalar gluonic cloud in a small Wilson loop in the one-instanton approximation
follows a similar reasoning with

\be
\left<F^2(x)F^2(x)\right>\approx \frac{(192)^2}{\pi^2}\int d\mu_{B+F} \frac{\rho^8}{ (|x-x_o|^2+\rho^2)^8}
\ee
so that

\be
\frac{\left< F^2(x)W\right>_c}{\left< F^2\right>\left<W\right>}\approx -\frac{384\pi^2}{7N_c}\frac{a^4}{\bar{\rho}^{4}}
\ee
Since $\left<F\tilde F F^2\right>=\left<q^\dagger \gamma_5 qF^2\right>=0$ by parity, the
pseudo-scalar condensates around a small Wilson loop vanish. Both the scalar quark and gluon condensates decrease near a small Wilson loop in 
the one-instanton approximation.

\section{Large Wilson loop in the QCD Instanton Vacuum}

The QCD instanton vacuum involves an ensemble of instantons and anti-instantons.
In this case, the one loop density (\ref{DENS}) is substituted by
 its mean value for the average instanton size~\cite{Shuryak:1978yk}

\be
\frac{d\mu_{B+F}}{d^4x d\rho}\rightarrow n\,\delta (\rho-\bar{\rho})
\ee
In particular,  the one-instanton zero mode in (29) turns to a quasi-zero mode with $im \rightarrow \lambda+im $ with a semi-circular spectral distribution $\nu(\lambda)$ (normalized to 1) so that
\be
<q^\dagger q>=n\int\frac{\nu(\lambda)d\lambda}{\lambda+im}\rightarrow-i\pi n\,\nu(0)
\label{QQCB}
\ee
in the chiral limit. (32) is the Banks-Casher relation~\cite{Banks:1979yr}
between the spectral density and the chiral condensate
in the QCD instanton vacuum. Note that in Euclidean
space, the chiral condensate is purely imaginary with the
convention $\bar q= (iq)^\dagger$.

Large Wilson loops in the QCD vacuum are screened by light quarks as shown in Fig.~\ref{WILLABX}.
The Wilson loop as a correlator $\left<{Q}^\dagger (a)Q(0)\right>$ can be thought as a heavy-light
meson correlator $\left<{Q}^\dagger\gamma q(a)\,{q}^\dagger \gamma Q(0)\right>$ after screening
with $\gamma=(1, \gamma_5)$. The almost near degeneracy of the scalar and pseudo-scalar
bound heavy-light states follows from a combination of heavy-quark and chiral symmetries~\cite{Nowak:2003ra,Bardeen:1993ae} and
was confirmed experimentally~\cite{Aubert:2003fg,Besson:2003cp}. In this section,
we will analyze both the fermionic and gluonic condensates in a large Wilson loop screened by
light quarks using the QCD instanton vacuum.

\subsection{Scalar Quark Condensate}

A measure of the scalar
condensate in the large Wilson loop of Fig.1 (right) amounts to the contributions shown in
Fig.~\ref{WILLCOND}. The screened Wilson lines in QCD are tied to the condensate insertion (black blob)
through a pair of re-scattering quarks in the QCD vacuum. The gray blobs are the heavy-light $Qqq$ effective
vertices. The contribution on the left is of order  $\kappa_{qqQ}$ while that on the right is of order $\kappa^2_{qqQ}$ and
sub-leading. The leading contribution is ($x_W=(\tau, {\bf 0})$)

\begin{eqnarray}
\frac{\left<q^\dagger q(x)W\right>_c}{\left<W\right>}
=2i\frac{\kappa_{qqQ}}{a_{qQ}^3} \int d\tau
\left<T^* q^\dagger q(x_W)  q^\dagger q(x)\right>
\label{CC}
\end{eqnarray}
 The factor of 2 is due to the two possible insertions on the screened heavy quarks.

From (\ref{C}) it follows that

\begin{eqnarray}
\frac{\left< q^\dagger q(x)W\right>_c}{\left<W\right>}\approx 2i\frac{\kappa_{qqQ}}{a_{qQ}^3}
\left(C^0(L)+C^1(L)\right)
\label{CCI}
\end{eqnarray}
with the probing point at $x=(x_4,L)$. Because of $x_4$-translational invariance, (\ref{CCI}) depends
only on the spatial distance $L$.
The purely imaginary nature of the result in Euclidean space is due to the zero modes.
Using (\ref{QQCB}) we can re-write (\ref{CCI}) in real form in Minkowski space as

\begin{eqnarray}
\label{QQW}
\frac{\left< \bar q q (x)W\right>_c}{\left<\bar q q\right>\left<W\right>}
\approx -\frac{2\kappa_{qqQ}}{a_{qQ}^3\pi n\nu(0)}
\left(C^0(L)+C^1(L)\right)
\end{eqnarray}

The connected contribution to the scalar quark correlator is

 \begin{eqnarray}
&&C^0(L)=\\
&&\frac{M_0^4N_c}{4\pi^4} \int d\tau \frac{K_2(M_0 \sqrt{L^2+\tau^2})^2-K_1(M_0 \sqrt{L^2+\tau^2})^2}{L^2+\tau^2}\nonumber
\end{eqnarray}
and asymptotes
\be
 C^0(L)\approx
 \frac{3N_cM_0^5}{8\pi^3} \frac{e^{-2M_0 L}}{(M_0L)^3}
\label{CC0}
\ee
after using the expansion
\be
K_{\alpha}(x)=\sqrt{\frac{\pi}{2x}}e^{-x}(1+\frac{2\alpha^2-1}{8x}+...)
\ee
The disconnected contribution follows similarly

\begin{eqnarray}
C^1(L)\approx
\frac{{g_\sigma^2}\,{{m_\sigma}}}{\sqrt {32\pi^3 }}\,\frac{e^{-m_\sigma L}}{\sqrt { m_\sigma L}}
\label{CCX}
\end{eqnarray}
Both (\ref{CC0}) and (\ref{CCX}) are positive making the result (\ref{QQW}) negative.
The scalar quark condensate depletes near a large Wilson loop in overall agreement 
with the lattice results~\cite{Iritani:2014jqa,Buerger:1993bq}



\subsection{Pseudo-scalar Quark Condensate}

\begin{figure}[htb]
\centerline{
\includegraphics[width=6cm]{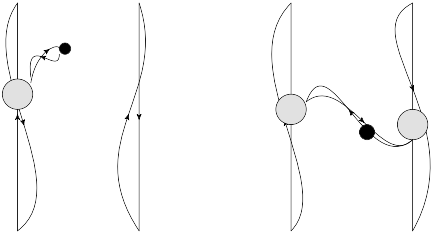}}
\caption{Screened large Wilson loop of width $a$ with a condensate insertion (black blob). See text.}
\label{WILLCOND}
\end{figure}

The pseudo-scalar condensate in a large and screened Wilson loop can be sought by similar arguments.
However, we note that the leading contribution of order $\kappa_{qqQ}$ as shown in Fig.~\ref{WILLCOND} (left)
involves the Fierzed pseudo-scalar vertex $u^\dagger \gamma_5 u d^\dagger\gamma_5 d$ from (\ref{qqQ}).  Let
$Q^\dagger \gamma u$ be a pair of heavy-light chiral partners with $\gamma=(1,\gamma_5)$. The induced vertex is

\begin{eqnarray}
\left< 0|u^\dagger\tilde\gamma Q\, \left[Q^\dagger(1+i\gamma_4)Q\,  u^\dagger\gamma_5 u\right]  Q^\dagger\gamma u|0\right>
\otimes d^\dagger\gamma_5 d
\label{VERTEX}
\end{eqnarray}
For $\tilde\gamma=\gamma$ (\ref{VERTEX}) vanishes by spin tracing. However, for nearly degenerate
chiral partners by heavy quark symmetry~\cite{Nowak:2003ra,Bardeen:1993ae}, the transition vertex (\ref{VERTEX}) with
$\tilde\gamma=(\gamma_5,1)$  does not vanish.  A rerun of the arguments for the scalar condensate gives

\begin{eqnarray}
\frac{\left<q^\dagger \gamma_5q(x)W\right>_c}{\left<W\right>}
=2i\frac{\kappa_{qqQ}}{a_{qQ}^3}
 \int d\tau \left<T^*q^\dagger \gamma_5q (x_W)q^\dagger\gamma_5 q (x)\right>
\label{CC}
\end{eqnarray}
From (\ref{C5}) it follows that

\begin{eqnarray}
\frac{\left<\bar q \gamma_5q(x)W\right>_c}{\left<W\right>}
\approx -2\frac{\kappa_{qqQ}}{a_{qQ}^3}\left(C_5^0(L)+C_5^1(L)\right)
\end{eqnarray}
after translation to Minkowski space with $\gamma_5^E=-i\gamma_5^M$ and $\bar q=(iq)^\dagger$.
The connected and disconnected contributions to the pseudo-scalar quark correlator are respectively,

\begin{eqnarray}
C^0_{5}(L) =\frac 13\, C^0(L)
\end{eqnarray}
and

\begin{eqnarray}
C_5^1(L)\approx \frac{g^2_\pi m_{\eta^\prime}}{\sqrt {8\pi^3 }}
\frac{e^{-m_{\eta^\prime}L}}{\sqrt{m_{\eta^\prime}L}}
\end{eqnarray}
which is to be compared to (\ref{CCX}).







\subsection{Scalar Gluon Condentate }

The gluon condensate surrounding a large Wilson loop follows similar arguments. First, we note that
the screened ${Q}^\dagger q$ state acts as a colored electric dipole of size $a_{qQ}$ sourcing
a composite Coulomb field. An estimate of the pertinent electric vertex follows from second order
perturbation theory~\cite{Shuryak:2000df,Liu:2014qrt}

\be
\frac{|\left<0|(a_{qQ} E^a)|1\right>|^2}{(g^2/a_{qQ})}\approx (a_{qQ}^3/g^2) E^aE^a
\label{VV}
\ee
where the perturbative Coulomb splitting between the singlet and octet state was used.
This approximation is valid  as long as the perturbative Coulomb potential is greater
than the instanton Coulomb potential in the bound state or $g/a_{qQ}\gg 1/g\bar \rho$.  In the
converse, then $g^2/a_{qQ}\rightarrow 1/\bar\rho$  so that

\be
\frac{|\left<0|(a_{qQ} E^a)|1\right>|^2}{(1/\bar\rho)}\approx (a_{qQ}^2\bar\rho) E^aE^a
\label{VVV}
\ee
For our problem $a_{qQ}\approx \bar\rho\approx 1/3$ fm and $g\equiv g(\bar\rho) \approx 2$ making (\ref{VV})
more appropriate for the screened Wilson loop.

With this in mind, a rerun of the
preceding arguments give

\begin{eqnarray}
\frac{\left<F^2(x)W\right>_c}{\left<W\right>}
=+2\frac{a_{qQ}^3}{g^2}\int\,d\tau\,\frac 12\,\left<T^*F^2(x_W)F^2(x)\right>
\end{eqnarray}
Instantons carry $E=B$ in Euclidean space by self-duality, i.e. $E^2\equiv F^2/2$.
So the gluonic insertion measures the amount of gluonic correlations

\begin{eqnarray}
\frac{\left<F^2(x)W\right>_c}{\left<W\right>}
=\frac{a_{qQ}^3}{g^2}\frac{N_f\sigma_\star^4 m_s}{2\pi^2f_\sigma^2}\int d\tau\frac{K_1(m_s\sqrt{L^2+\tau^2})}{\sqrt{L^2+\tau^2}}
\end{eqnarray}
For large $L $,

\begin{eqnarray}
\label{FFW}
\frac{\left<F^2(x)W\right>_c}{\left<F^2\right>\left<W\right>}
 \approx +\frac {N_f}{64\pi^2} \frac {a_{qQ}^3}{g^2}  \frac{\sigma_\star^4{m_s}}{nf_\sigma^2}\frac{e^{-m_s L}}{\sqrt {2\pi^3 m_sL}}
\end{eqnarray}
after using $\left<F^2\right>=32\pi^2n$. The gluon condensate is enhanced near a large Wilson loop.

\subsection{Pseudo-scalar Gluon Condensate}

The pseudo-scalar gluon condensate  involves the transition vertex discussed in (\ref{VERTEX})
for nearly degenerate chiral heavy-light states, and mixes $F\tilde F$ with $q^\dagger\gamma_5 q$.
Specifically,

\begin{eqnarray}
\frac{\left<F\tilde F (x)W\right>_c}{\left<W\right>}
=2i\frac{\kappa_{qqQ}}{a_{qQ}^3}
 \int d\tau \left<T^*q^\dagger \gamma_5q (x_W) F\tilde F (x)\right>
\label{CCFF}
\end{eqnarray}
which can be reduced using

\begin{eqnarray}
 \left<T^*q^\dagger \gamma_5q (x) F\tilde F (0)\right>=
 -4i\frac{g_\pi\chi_*m_{\eta^\prime}}{N_cf^2_\pi}\frac{K_1(m_{\eta^\prime}|x|)}{|x|}
\end{eqnarray}
Setting $x=(x_4,L)$ and using $x_4$-translational invariance we obtain

\begin{eqnarray}
\frac{\left<F\tilde F (x)W\right>_c}{\left<W\right>}=
+\frac{\kappa_{qqQ}}{a_{qQ}^3} \frac{2{\pi} g_\pi}{N_c}\frac {\chi_*m_{\eta^\prime}}{f_\pi^2}
\frac{e^{-m_{\eta^\prime}L}}{\sqrt{\pi m_{\eta^\prime}L}}
\end{eqnarray}
Since $g_\pi\equiv \left<\bar q q\right>/f_\pi <0$, we conclude that the topological density decreases 
near a largeWilson loop.

\subsection{Squared Scalar and Pseudo-scalar Gluon Condensates}

The squared scalar and pseudo-scalar gluon condensates around a fixed and large Wilson loop can also
be analyzed using similar arguments. The squared pseudo-scalar condensate reads

\begin{eqnarray}
\frac{\left<(F\tilde F (x))^2 W\right>_c}{\left<W\right>}
=2i\frac{\kappa_{qqQ}}{a_{qQ}^3}
 \int d\tau \left<T^*q^\dagger q (x_W) (F\tilde F)^2 (x)\right>
 \label{FFT2}
\end{eqnarray}
The correlation function in (\ref{FFT2}) can be obtained by extending the analysis
in~\cite{Kacir:1996qn}. The result is

\begin{eqnarray}
 \left<T^*q^\dagger q (x) (F\tilde F)^2 (0)\right>=2i{\cal B}\,\left(\frac{16\chi_\star m_{\eta'}}{f_\pi^2}\frac{K_1(m_{\eta^\prime} |x|)}{|x|}\right)^2
\end{eqnarray}
The overall constant ${\cal B}$ is set by the momentum dependent constituent mass (\ref{MK})

\begin{equation}
{\cal B}=\int \frac{d^4 k}{(2\pi)^4} \frac{M(k) (k^2-M^2(k))}{(k^2+M^2(k))^2} 
\end{equation}
which is numerically positive: ${\cal B}\approx (150\,{\rm MeV})^3$. 
Thus
\begin{eqnarray}
\frac{\left<(F\tilde F (x))^2 W\right>_c}{\left<(F\tilde F)^2\right>\left<W\right>}
\approx -\frac{\kappa_{qqQ}}{a_{qQ}^3}\,\frac{256\pi{\cal B} }{\chi_*/V_4}\frac{ \chi^2_\star m^3_{\eta'}}{f^4_\pi}\left(\frac{e^{-m_{\eta^\prime}L}}{m_{\eta^\prime}L}\right)^2
\label{FFT22}
\end{eqnarray}
where the normalization 

$$\left<(F\tilde F)^2\right>\approx \chi_*\delta^4(0)\equiv \frac{\chi_*}{V_4}$$ 
following from (\ref{FFTFFT}) was used. (\ref{FFT22}) is overall negative, an indication that 
the squared topological density is slightly depleted near a large Wilson loop.
In the quenched approximation the analogue of (\ref{FFT22}) was analyzed on the lattice~\cite{Faber:1993fj} with the result that the
squared topological density decreases near a static color source described by a Polyakov line.

The same reasoning for the squared gluon condensate yields

\begin{eqnarray}
\frac{\left<(FF (x))^2 W\right>_c}{\left<(FF)^2\right>\left<W\right>}
\approx \frac{\kappa_{qqQ}}{a_{qQ}^3}\,\frac{256\pi{\cal C}}{(32\pi^2n)^2}\,\frac{  \sigma^4_\star m^3_s}{f^4_\sigma}\left(\frac{e^{-m_sL}}{m_sL}\right)^2
\end{eqnarray}
with the constant ${\cal C}$ given by the running constituent mass (\ref{MK})

\be
{\cal C}=\int \frac{d^4 k}{(2\pi)^4}\frac{M^3(k)(3k^2-M^2(k))}{(k^2+M^2(k))^3}
\ee
Numerically: ${\cal C}\approx (207\,{\rm MeV})^3$,  so that the squared gluon condensate is slightly enhanced near a large Wilson loop.


\section{Dipole-Dipole Interaction in the QCD Instanton Vacuum}

\begin{figure}[htb]
\centerline{
\includegraphics[width=6cm]{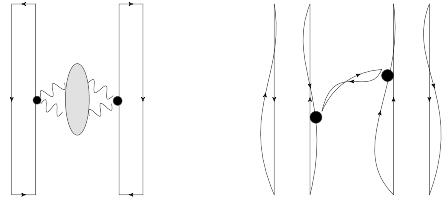}}
\caption{Static dipole-dipole correlator: small size and unscreened dipoles (left) and large size and screened dipoles (right). See text.}
\label{WILLABXX}
\end{figure}

The dipole-dipole interaction in the QCD instanton vacuum receives contributions from
the gluon-gluon corralator for $a\leq a_{qQ}$ and from scalar correlations for $a\gg a_{qQ}$
as shown in Fig.~\ref{WILLABXX}. They will be addressed separately below. Generically, the dipole-dipole
potential follows from

\begin{equation}
V(L)=-\lim_{T \to+\infty} \frac {1}{T}\,\,
 {\rm ln} \left(\frac{\left<W(L)W(0)\right>_c}{\left<W(L)\right>\left<W(0)\right>}\right)
 \label{DD}
\end{equation}
with $W(0)$ and $W(L)$  two identical and rectangular Wilson loops of
width $a$ and infinite time-extent $T$, centered at $0$ and $L$ respectively.
Throughout this section $L$ refers to the spatial separation between the two
dipoles. For simplicity, the dipoles have identical width and are parallel.

\subsection{Gluon Induced Correlations: $a\leq a_{qQ}$}

For small dipole sizes $a\leq a_{qQ}$, the effects of screening can be ignored,
and the dipoles source (colored) electric fields as depicted in Fig.~\ref{WILLABXX} (left). As we noted earlier, the dipole sources
a composite vertex $(a^3/g^2) E^2$ for $g/a>1/g\bar \rho$ and $(a^2\bar \rho) E^2$ for $g/a<1/g\bar \rho$.
Recall that $gE\equiv E$ in our notations. For the former

\begin{eqnarray}
V_G(L)\approx -\frac{a^6}{g^4}\int\,d\tau\,\frac 14\,\left<T^*F^2(\tau,0)F^2(0,L)\right>
\end{eqnarray}
while for the latter $a^6/g^4\rightarrow a^4{\bar \rho}^2$.
Using the above arguments for the scalar gluon correlator in the QCD instanton vacuum we obtain
(\ref{GLUON})

\begin{eqnarray}
V_G(L)=-\frac{a^6}{g^4}
\frac{N_f\sigma_\star^4m_s}{8\pi^2f_\sigma^2}\int d\tau\frac{K_1(m_s\sqrt{L^2+\tau^2})}{\sqrt{L^2+\tau^2}}
\end{eqnarray}
or asymptotically

\begin{eqnarray}
V_G(L)\approx  -\frac{a^6}{8g^4}\frac{N_f\sigma_\star^4 {m_s}}{3f_\sigma^2}\frac{e^{-m_s L}}{\sqrt {{2\pi^3}m_sL}}
\label{VG}
\end{eqnarray}
\newline
The dipole-dipole potential is screened by the light scalar mass (\ref{GLUON})

$$m_s=\left(m_\sigma^2-2N_f\sigma_*^2/f_\sigma^2\right)^{1/2}\approx 349\,{\rm MeV}$$
which is smaller than the sigma meson mass $m_\sigma\approx 640\,{\rm MeV}$ due to the finite compressibility 
$\sigma_*^2\approx 0.36/{\rm fm}^4$ of the QCD instanton vacuum. As we noted in section IIB, the effect of
the compressibility on the scalar mass $m_s$ are of order $N_f/N_c^2$.

The alternative regime with $a^6/g^4\rightarrow a^4{\bar \rho}^2$ is to be compared to the single instanton result
of $(a^4{\bar \rho}^2)/L^7$ in~\cite{Shuryak:2000df}.
Non-paralell dipoles yield additional contributions to $V_G(L)$ depending on the dipoles relative
orientations. They will not be discussed here.

\subsection{Fermion Induced Correlations: $a\gg a_{qQ}$}

Large size dipoles with $a\gg a_{qQ}$ are screened by the light quarks as depicted in Fig.~\ref{WILLABXX} (right). They form four pairs of heavy-light bound states. Their pair interactions are then similar to the ones we discussed before, with

\begin{eqnarray}
V_0(L)
\approx -\frac{\kappa_{qqQ}^2}{a_{qQ}^6}\frac{3N_cM_0^5}{4\pi^3}\frac{e^{-2M_0L}}{(M_0L)^3}
\end{eqnarray}
for the connected or $C^0$ contribution, and

\begin{eqnarray}
V_1(L)\approx -\frac{\kappa_{qqQ}^2}{a_{qQ}^6}
\frac{g^2_\sigma {m_\sigma} }{64}\frac{e^{-m_\sigma L}}{\sqrt {\pi^3 m_\sigma L}}
\label{V1}
\end{eqnarray}
for the disconnected or $C^1$ contribution.
All pair contributions add to the total fermionic dipole-dipole interaction

\be
V_{F}(L)=\sum_{i=0,1}\left(V_i(L)+V_i(L+2a)-2V_i(L+a)\right)
\ee
We also note that the leading fermionic induced interaction following by Taylor expansion

\be
V_F(L)\approx a^2 \frac{\partial^2}{\partial^2 L}\left( V_0(L)+V_1(L)\right)
\ee
is also attractive. Since $m_s<m_\sigma$, the gluonic induced interaction (\ref{VG}) is dominant
over the fermionic induced interaction (\ref{V1})  for dipoles of sizes $a\gg a_{qQ}\approx 0.26\,{\rm fm}$.


\section{Conclusions}

The QCD instanton vacuum offers a semi-classical framework for discussing quark and gluon
correlations. It breaks explicitly conformal symmetry through the instanton density, and spontaneously
chiral symmetry through the appearance of a chiral condensate. As a result a light pseudo-scalar
flavor octet of Goldstone mesons emerge which satisfies the GOR relation. A remaining flavor singlet
meson is turned heavy through the finite topological susceptibility of the QCD instanton vacuum
in accordance with the U(1) axial anomaly in QCD. Scalar and pseudo-scalar gluon
correlators  in the QCD
instanton vacuum mix with the scalar and pseudo-scalar light meson corelators.

Large Wilson loops are screened in the QCD instanton vacuum. We have explicitly analyzed the
scalar and pseudo-scalar quark and gluon content of large Wilson loops and shown that they are
dominated by the light scalar and pseudo-scalar meson exchanges at large distances. Because
of the lack of confinement in this model, the intermediate distances are contaminated by the two
constituent quark exchange as well. Recall that the latter is dominant at shorter distances.

Recently, we have analyzed similar issues in the context of AdS/QCD~\cite{Liu:2014qrt},
 in the double limit of a large number of colors $N_c$ and strong $^\prime$t Hooft
coupling. In leading order in $1/N_c$, large Wilson loops are unscreened. In contrast, our
results are derived at weak coupling for which the semi-classical of the QCD instanton
vacuum is justified and fixed $N_c$ for which screening is important since in our case $N_f/N_c=1$.
Lattice QCD simulations support the depletion of the scalar quark condensate in the vicinity
of a Polyakov line~\cite{Buerger:1993bq} and a Wilson loop~\cite{Iritani:2014jqa}.

We have explicitly analyzed small and large static dipole-dipole interactions in the screened QCD instanton
vacuum. Our analysis shows that these interactions are attractive, short ranged and dominated by the exchange of
a light scalar $m_s<m_\sigma$. Owing to screening, the scalar  glueballs mix in the QCD instanton vacuum, causing
the static dipole-dipole interaction to deviate from the Casimir-Polder form noted in the one-instanton approximation~\cite{Shuryak:2000df}.

Static dipole-dipole interations provide a simple estimate of the
static interaction between QCD strings. These interactions play an important role in addressing issues of interactions
of QCD strings, in particular near the Hagedorn temperature where they may condition the nature of the transition
to a fireball as a stringy black-hole~\cite{Shuryak:2013sra,Kalaydzhyan:2014zqa}.  We hope to address some related issues next.

\section{Acknowledgements}

This work was supported by the U.S. Department of Energy under Contracts No.
DE-FG-88ER40388.

 \vfil

\end{document}